\documentstyle[12pt]{article}
\begin{document}


{\it \hspace*{\fill} DOE/ER/40762-073\\
\hspace*{\fill} U. of MD 96-057} 

\vspace*{.25in}

\noindent { \large \bf \boldmath Large-$N_c$ and Chiral Limits of QCD and
Models of the Baryon}
\vspace*{.12in}

\hspace*{.4in}Thomas D. Cohen
\vspace*{.12in}

\hspace*{.4in}{\it Department of Physics, University of~Maryland,
College~Park, Maryland~20742}

\vspace*{.25in}

\noindent Two ideas have greatly contributed to our understanding
of baryon structure in the framework of Quantum Chromodynamics (QCD).
The first, chiral symmetry,
received its fundamental justification from 
QCD and has been developed into the powerful technique of
chiral perturbation theory.  The other notion is the large-$N_c$ limit,
namely the behavior of QCD in the limit where there are a large
number of colors.  Like chiral symmetry, the large-$N_c$ limit predicts
 many relations between baryon
matrix elements.  These so-called ``model-independent'' predictions
should agree with each other, but paradoxically that is not
always the case. The paradox is resolved by recognizing that the
two limiting processes (large-$N_c$ and the chiral limit) do 
not commute.  This noncommutivity  can be traced to the  special role
played by the $\Delta$ resonance in large-$N_c$ QCD. This gives rise to
a new parameter in the effective
theory of baryons. 

\vspace{.25in}
\noindent CONTENTS\\ 
\hspace*{.04in}I.~Introduction\\
\hspace*{.02in}II.~The Skyrme Model, Its Hedgehog Relatives, and a Chiral Puzzle\\
III.~Baryons in the Large $N_c$ and Chiral Limits \\
\hspace*{.01in}IV.~Implications for Modeling Baryons\\
Acknowledgments\\
Appendix A: Chiral Properties in  Skyrme Models\\
Appendix B: Consistency and Large  $N_c$\\
References\\

\newpage

\section{INTRODUCTION}

It has been known since the discovery of the large anomalous moment of
the proton in the 1930s that the proton is not a fundamental Dirac
particle; it has an internal structure.  Since this
time, understanding the structure of the nucleon--and of other
baryons--has been an  important problem in theoretical physics.  
Early attempts to model the  structure of the baryon were based on
a cloud of virtual mesons.  In the 1960s an entirely
different approach was developed in which the structure of the nucleon and
other baryons was explained in terms of the quark model.  For the past
two decades, it has been understood that the
nucleon is a stable state in quantum chromodynamics (QCD), the
gauge theory underlying the strong interactions.   Thus the
structure of the nucleon is understood as arising from the interactions 
of quarks and gluons in QCD.

Why is the  problem of baryon structure still of interest today?
One primary reason is that we cannot directly solve QCD in
the low energy regime and hence we cannot directly compute
baryon properties from the underlying theory.  Thus, at present, the
theoretical study of baryons depends largely on phenomenological models
and on exploiting those features of QCD which are tractable.
Recently, significant progress has been made in understanding some of
these tractable features of QCD including aspects of the
$1/N_c$ expansion (where $N_c$ is the number of colors) and on the
chiral expansion.  This colloquium is intended to review 
some of these developments with a particular emphasis on the
implications for the various models of baryon structure that are in
use. 

The models discussed here are not designed to  describe accurately all
aspects of nucleon structure.  In particular, they are meant to describe 
  low-momentum-transfer observables such as the magnetic moment, mean
square charge and magentic radii, and $g_A$, the axial vector current
coupling constant.  More generally, they should
describe the low momentum   form factors and transition form factors
for the various currents which can couple to baryons.  The models are
{\it not} designed to predict high-momentum-transfer
observables such as these form factors at large $Q^2$ or the structure
functions which parameterize deep inelastic scattering.

It might be argued that eventually these considerations will become irrelevant.
In time, numerical simulations of lattice QCD may become good
enough so that reliable model-independent predictions of baryon
properties may be obtained.\footnote{Present lattice QCD simulations
of hadronic properties
have significant systematic uncertainties.   The present state of the art is
nicely summed up in Karsch {\it et al.} (1995).}  However, while high
quality numerical simulations may allow us to test whether QCD can explain
low-energy hadronic phenomena, they will not, by themselves, give
much insight into how QCD works in the low-energy regime.  
Simple intuitive pictures are essential to obtain insight, and models
provide such pictures.   Condensed matter physics provides a useful
analogy: even if one were able to solve the electron-ion many-body
Schr\"{o}dinger  equation by brute force on a computer
and directly predict observables, to have any real understanding of
what is happening, one needs to understand the  effective degrees
of freedom which  dominate the physics such as phonons, Cooper pairs,
quasiparticles, and so forth.   To gain intuition about 
these effective degrees of freedom, modeling is required.   In much
the same way, models of the hadrons are essential in developing
intuition into how QCD functions in the low energy domain.

Of course, there are a plethora of models of the nucleon on the 
market.   Most of these models seem, at least superficially, to
be quite different.   As two extreme examples consider
the nonrelativistic constituent quark model (for an introduction to
this model see Close (1979); Isgur (1992); Karl (1992)) and the Skyrme
model (Skyrme, 1961a, 1961b, 1962; there are many reviews of the Skyrme
model--see for example Zahed and Brown, 1986).  The degrees of
freedom seem to be completely different.  In one case the nucleon
is thought of as a bound state of three constituent quarks, and in the
other as a topological soliton comprised entirely of meson
fields.   (A baryon in the Skyrme model is topologically stable in 
much the same way that a knot in an infinite piece of string is
topologically stable; one cannot ``untie'' the
meson field configuration of a topological soliton by continuously 
changing the fields.)

As will be seen in the course of this colloquium, despite the apparent
differences, these models share important similarities, some of which
are not completely obvious.  Accordingly, when
assessing the models it is important to distinguish features which
are peculiar to the model and those arising from general features of
QCD.  An obvious,  trivial example of a general feature which
constrains models is isospin symmetry (or, more generally,  flavor symmetry).
Any  sensible model is isospin-symmetric to leading order
and will predict a degenerate proton
and neutron.  The fact that a particular model predicts this degeneracy should not 
be viewed as a triumph of the model, but merely as a necessary condition
on sensible model building.  

In much the same way, the large-$N_c$ behavior of any sensible model of
baryons is strongly constrained by QCD (Witten, 1979).  Moreover,  this
behavior, particularly the spin-flavor structure (Gervais and Sakata,
1984a, 1984b; Dashen and Manohar, 1993; Dashen {\it et. al}, 1994, 1995) 
is highly nontrivial
and one finds model-independent relations between nucleon and
$\Delta$ properties which become exact as $N_c$ gets large.   (The
$\Delta$ is the lowest lying
non-strange baryon resonance.  It has  a spin of 3/2 and isospin of 3/2.)
 For example, the $\pi$-N-$\Delta$ coupling constant, $g_{\pi N\Delta}$
 is predicted to be 3/2 times $g_{\pi N N}$, the
pion-nucleon coupling constant.
These
relations may all be summarized in the statement: as $N_c \rightarrow
\infty$, the low lying states in the system obeys a SU$(2n_f)$ symmetry, with  $n_f$
is the number of light flavors.  This is just the symmetry of
the nonrelativistic  quark model with $n_f$ quarks, each having
two spin orientations.   The fact that the large-$N_c$ behavior
of the model must match QCD constrains the predictions of the model up
to order $1/N_c$ type corrections.  In this sense,  although Skyrme models 
and constituent quark models are quite different,  both classes  of
models are constrained to have certain relations among observables
(up to $1/N_c$ corrections) ; as far as the relations between
these observables are concerned, the models are similar.
Of course, in the real world $N_c=3$ so that $1/N_c$ corrections may be
large--naively one would expect they are typically of relative order
1/3.   However, recently it has been observed that, for certain 
relationships (such as for isovector axial or magnetic
couplings) these relations must  hold to order $1/N_c^2$ (Dashen and
Manohar, 1993; Dashen {\it et al.}, 1994, 1995)  and thus are naively at
the level of about 10\%.

QCD also has approximate chiral symmetry\footnote{Chiral symmetry is
defined as the invariance of the theory under the transformation $q
\rightarrow {\rm e}^{i \vec{\alpha} \cdot \vec{\tau} \gamma_5} q$ where
$q$ is the quark field, $\vec{tau}$ is an isospin matrix, $\gamma_5$ is
a Dirac matrix and $\alpha$ is an arbitrary vector.  This
transformation mixes the upper and lower components
of the Dirac spinors.  The only term in the QCD Lagrangian which
violates this symmetry is the quark mass term.}
 due to the fact the
quark masses in the QCD Lagrangian (the``current quark masses'')
are much less than the typical
hadronic mass scale.   As is well known, apart from the explicit symmetry
breaking due to the nonzero quark masses, there is also spontaneous
symmetry breaking; the ground state of the theory, the vacuum, is not
invariant under chiral rotations even though
the Lagrangian is.   This has profound consequences: Goldstone's
theorem tells us that in the absence of explicit symmetry breaking
there must be massless particles--particles which are identified with the
pions (or more generally, with the pseudoscalar octet).   Moreover, 
neglecting finite quark mass effects, the interactions of pions with
each other and with baryons are also constrained by the symmetry to
vanish at zero momentum transfer. (The constraints of chiral symmetry
are nicely described in the recent review by  Bernard, Kaiser and
Meissner (1995).)  It is quite sensible to treat 
the light quark masses  as a perturbation 
and develop a systematic expansion.  This expansion
is chiral  perturbation theory ($\chi PT$).   The natural parameter is
generally taken to be $m_\pi^2/\Lambda^2$, where 
$m_\pi^2$ is proportional to $m_q$ (due to the Gell-Mann--Oakes--Renner
relation (Gell-Mann {\it et al.}, 1968)) and $\Lambda$ is a typical 
hadronic mass scale parameter, such as $M_N$, $m_\rho$ or $4 \pi f_\pi$.

 Chiral symmetry strongly constrains certain properties of the
nucleon and its interactions with pions, particularly  $\pi$-N scattering,
 and one might wish to view 
chiral properties as providing a necessary constraint on a reasonable
model of the nucleon.  Many models such as Skyrme models, chiral or
hybrid bag models (reviewed in Vepstas and Jackson, 1990), chiral quark
meson models (reviewed in Banerjee, Broniowski and Cohen, 1987; Birse, 
1991)   and baryon models based on the Nambu--Jona-Lasinio model
(reviewed in  Alkofer,  Reinhardt and Weigel,  1995)    
  are explicitly chirally symmetric. On the other hand,
many models of baryons do not respect chiral symmetry.  Famous examples
include both the constituent quark model and the MIT bag model (Chodos
{\it et al.}, 1974).
One attitude that can be taken is that as long as one avoids studying
physics which is sensitive to low momentum
$\pi$-N interactions, such as $\pi$-N scattering,  the explicit role
of chiral symmetry is small and  its indirect effects may
be simulated by a good non-chiral phenomenology.  

There  are certain interesting subtleties involving the large-$N_c$ and
chiral expansions.  For example, in the combined large-$N_c$ and chiral limits there
are new model-independent predictions.  These predictions can be 
derived from pion loop effects in large-$N_c$ chiral perturbation
theory--a generalization of usual chiral perturbation theory which
accounts for the large-$N_c$ behavior of the quantities involved (Cohen
and Broniowski, 1992).  
If one looks at these predictions one immediately observes that
 the large-$N_c$
and chiral limits do not always commute; for a number of quantities
the qualitative result depends on whether one first takes the chiral or
large-$N_c $ limit.  
If such a quantity is treated as a function of $m_q$, then the $1/N_c$
expansion is not uniformly convergent; conversely, if  such a
quantity is treated as a function of $N_c$, then the chiral expansion
is not uniformly convergent.  The origin of this behavior is
ultimately rather easy to trace.  In the large-$N_c$ limit the
N-$\Delta$ mass splitting goes as $1/N_c$. This is seen explicitly in
the Skyrme model (Adkins, Nappi and Witten, 1983) and can be derived in
general  from large-$N_c$ consistency conditions (Jenkins, 1993). 
Consider the ratio between the pion mass and the N-$\Delta$ mass splitting:

\begin{equation}
d \equiv \frac{M_\Delta -M_N}{m_\pi} \; \; .
\label{def:d}
\end{equation}

\noindent Clearly this ratio has the property that as one approaches the chiral
limit of $m_\pi \rightarrow 0$, then $d$ diverges; while if one
approaches the large-$N_c$ limit, then $d \rightarrow 0$.  
Quantities which are sensitve to the value of $d$ can have radically
different results in the two limits. 
 It is not surprising that quantities
could be highly sensitive to $d$.   As will be discussed later, 
loop  graphs with a virtual pion and an intermediate $\Delta$   produce
such effects.

 The noncommutativity of the large-$N_c$ and chiral  limits  for
certain observables raises some practical questions.  If the two
expansions are incompatible with each other for important quantities,
which expansion ought one use to describe the real world?    The
 ratio $d$ defined in Eq. (\ref{def:d}) is empirically
approximately 2.1; {\it i.e.}  neither large  nor small.  This suggests
that neither the chiral nor $1/N_c$ expansions will work well for
quantites which are sensitive to $d$.  It is far more natural to consider a combined
expansion in which $1/N_c$ and $m_\pi/\Lambda$ are both treated as
small but in which $N_c m_\pi/\Lambda$ is treated as order 1.  While
this expansion may be the natural one, most of the models in use  are
 not based on this expansion.    

In this colloquium the relationship of chiral symmetry, large-$N_c$ QCD,
and various models of the nucleon will be discussed.  The emphasis
will be on models such as the Skyrme model which attempt to build in
both the correct large-$N_c$ and chiral physics.  This colloquium is not 
intended as a comprehensive review of all facets of the problem; 
rather it should serve as a guide to some of the more intriguing
connections between various approaches to the problem of baryon
structure.  The discussions here will concentrate almost entirely on
the non-strange sector.
The reason for this is that most of the interesting issues of principle
are present for SU(2) flavor.  The extension to SU(3) flavor
is relatively straightforward in principle, but is  less
transparent--particularly when one considers the large-$N_c$ behavior.   

\section{THE SKYRME MODEL,  ITS HEDGEHOG RELATIVES, AND A CHIRAL PUZZLE}

This section will introduce an  apparent paradox concerning the Skyrme
model and other
models which are based on large- $N_c$ QCD
and approximate chiral symmetry.   Such models include the chiral-quark
meson soliton model,
 the chiral or hybrid bag model,
and the soliton approach
 to the Nambu--Jona-Lasinio (NJL)  model.  All of these models have a
few  important features in common and all of the issues discussed
concern  these models
in precisely the  same way as they concern the Skyrmion.

 The key features of these models are  that they are all based on a
mean-field theory  treatment (which is ultimately justified by an
appeal to large-$N_c$ physics) and that all of these models have the
long-range physics
dominated by a pionic tail.  It is worth stressing that all of the
physics which will be discussed depends only on large-$N_c$ physics and the
role of the pion tail.  Nothing in the following discussion depends
explicitly on the
fact that the Skyrme model is topological in nature.  Thus, the various
nontopological models  with the baryon number carried by
explicit quarks behave in the same way as the Skyrme model.

In the discussion in this section, I will restrict my attention to the
traditional treatment of these models in which one works systematically
in a $1/N_c$ expansion and typically only retains the leading order
term.  The lowest energy configuration in the mean-field
theory is static and the mean-field 
treatment requires that the pion field be classical.  As it happens, in
all of these models the lowest-lying  classical solutions break both
the rotational and isorotational symmetry; the minium energy
configurations are so-called hedgehogs with the pion's isospin
direction correlated with the spatial direction (Skyrme 1961a, 1961b,
1962).  Thus the pion field
assumes the form
\begin{equation}
{\vec \pi}_a({\bf r}) = \pi (r) \hat{r}_a \label{pion}
\end{equation}
where the  subscript $a$ in $\pi_a$ indicates the isospin component.  This form is
referred to as a hedgehog since the isospin points radially outward like
a hedgehog protecting itself.  If one rotates a hedgehog  in
either space or isospin space one gets a new configuration
which is degenerate in energy with the original hedgehog.
 
Of course physical states have good angular momentum and isospin. 
Thus, it is clear that the hedgehog or rotated hedgehog
configurations do not immediately  correspond to the physical states. 
What then are they? To find the properties of individual states it is
necessary to project onto states of good quantum numbers.
A  semiclassical technique has been developed to do this  
which should become increasingly valid in the large-$N_c$ limit.  The
technique involves the study of slowly rotating self-consistency
classical configurations and then quantizing the parameters which
specify the rotations (Adkins, Nappi and Witten, 1983; Cohen and
Broniowski, 1986).  The physical states are  baryon with $I=J$.     The
energy splittings in the band follow a rotational spectrum:
\begin{equation}
 E_J \, = \, E_0 \, + \,  \frac{J(J+1)}{2 \cal{I}} \label{rot}
\end{equation}
Here $\cal{I}$, the moment of inertia, and $E_0$ are both of order
$N_c$.  Thus the splitting of the low-lying states in the band is of 
order $1/N_c^2$ relative to the total mass.
The two lowest states in the band are identified as the nucleon and
the delta.  Predicted higher states in the band are generally assumed
to be artifacts of the $1/N_c$ approximation.  
Moreover,  the nucleon
and the delta (and these exotic high spin-isospin states) all  have an
identical  intrinsic state.  Thus many observables are related 
(at least up to $1/N_c$ corrections).

The states in the band are labeled by the quantum numbers $I=J$, $m$, and
$m_I$. 
Let us consider arbitrary operators, denoted by $A,B,C$
and $D$ which carry
isospin $i$ and angular momentum $j$, with third components $\mu_i$  and $\mu$.
  The following relations among matrix elements are true:
\begin{eqnarray}
\langle I,m,m_I |  A^{i=0,j=0} | I,m,m_I \rangle \hspace*{.3in} & = & a \nonumber \\
\langle I, m ,m_I |  B^{i=1,j=0}_{\mu_i=0} | I, m, m_I \rangle \hspace*{.3in} & = &
 b \, \,   m_I \nonumber \\
\langle I, m, m_I |  C^{i=0,j=1}_{\mu=0} | I, m, m_I \rangle \hspace*{.3in} & = &
c \, \,  m \nonumber \\
\langle I^\prime, m^\prime, m_I^\prime |  D^{i=1,j=1}_{\mu, \mu_I} | I,
m, m_I, \rangle \hspace*{.3in} & =  &   d  \, \,  X^{\mu^\prime
\mu}_{(I^\prime, m^\prime m_I^\prime) (I ,m, m_I)}
\label{mi} 
\end{eqnarray}
where the lower case coefficients $a, b, c, d$ are constants which
do not depend on the particular state $| I, m, m_I \rangle$ in
the band, 
and 
\begin{equation}
  X^{\mu^\prime
\mu}_{(I^\prime, m^\prime m_I^\prime) (I ,m, m_I)} \, \equiv \, 
\sqrt{\frac{2 I^\prime +1}{2 I+1}} \, \left ( \begin{array}{ccc} I & 1&
I^\prime \\  m_I  & \mu_i & m_I^\prime \end{array} \right ) \, \left (
\begin{array}{ccc} I & 1& I^\prime \\  m & \mu & m^\prime \end{array} \right )
\end{equation}

These relationships were recognized much earlier to be ``model
independent'' in  the sense that  all Skyrme models, and other
large-$N_c$ hedgehog models will satisfy these relations.  In fact, these
relations  can be tested for a number of observables and  they
appear to work quite well (Adkins and Nappi, 1985).
However, these results are model-independent in a much deeper 
sense--consistency of the large-$N_c$
expansion requires them to be true (Gervais and Sakita, 1984a, 1984b;  
Dashen and Manohar 1993; Jenkins, 1993;  Dashen, Jenkins, and  Manohar
1994, 1995; Broniowski, 1994).  This large-$N_c$
consistency argument is outlined in Appendix B. 

Of course,  while ratios between various observables can be fixed from
Eqs. (\ref{mi}), the actual values of the coefficients on the right side of
Eqs. (\ref{mi})  depend in detail on the dynamics of the model.  
In general,
for arbitrary operators with real world parameters
 one must simply calculate and
see how well the models do.  However, as was noted previously, the
models are designed with chiral symmetry in mind.
The models have a  well-defined chiral limit corresponding to
vanishing quark mass,  or equivently $m_\pi^2$ vanishing. 
It is interesting
to see how predictions of the model behave as one approaches this limit. 

 The pion is the lightest degree of freedom in the
problem.  Thus, the classical pion field of Eq. (\ref{pion})  at long
distance behaves like a p-wave Yukawa field with the radial dependence:
\begin{equation}
\hspace*{-2.2in} \vec{\pi}_a({\bf r}) \rightarrow \vec{\pi}^{\rm
asympt}_a({\bf r}) = \hat{{\bf r}}_a \,
 \frac{3 g_{\pi N N}}{8 \pi M_N} \, (m_\pi +
 1/r) \, \frac{{\rm e}^{-m_\pi r}}{r}.
\label{yuk}
\end{equation}
Here $g_{\pi N N}$ is the $\pi$-N coupling constant as determined in the
model,  and the factor of 3/8 emerges from the details of the
semiclassical projection. Some observables depend strongly on the
tail of the configuration.  If  the tail is weighted strongly  enough
for some quantity
 then its predicted value  will either diverge as $m_\pi
\rightarrow 0$ or go as $m_\pi$.  In such cases, one can obtain essentially
model-independent predictions for the behavior of the observables as one
approaches the chiral limit.   A few of these predictions are listed in
Table 1. The derivation of  these relations is 
semi-classical in nature; the derivation of one of these quantities is 
outlined in  Appendix A.
These predictions depend only on 
$m_\pi$,  and $g_{\pi N N}/M_N$; the vector-isovector observables also
depend on the $\Delta$ nucleon mass splitting.  All of these
predictions are ``model independent'' in the sense mentioned
above: they apply to any large-$N_c$ hedgehog model without regard to
the details of the dynamics.  There is implicit model dependence in the
actual predictions since the relations depend on $g_{\pi N N}/M_N$
(and perhaps on  $M_\Delta -M_N$) as given by the models, and the models
are not guaranteed to produce the correct value for   $g_{\pi N N}/M_N$
or $M_\Delta -M_N$.
Moreover,  these model-independent  predictions only apply to the
leading chiral nonanalytic part of the expressions for observables.  In
addition, there are chirally suppressed contributions which depend
explicitly on the details of the model;  for  realistic values of
$m_\pi$ these chirally suppressed contributions may be numerically
significant for some observables.   

Now we come to the puzzle:  One can evaluate all of these quantities
using chiral perturbation theory ($\chi$PT).    The  approach is
reviewed in Bernard, Kaiser and Meissner (1995).  The leading chirally
singular piece is determined entirely from pion one loop graphs and the
coefficients are fixed at lowest order in the theory.  The  strong
belief is that chiral perturbation theory should exactly reproduce QCD
in the limit as $m_\pi^2 \rightarrow 0$.  Thus, the $\chi$PT
calculation of the various 
chirally singular properties in Table 1 should be exact.
However,  in looking at Table 1, one sees that all of these $\chi$PT
predictions disagree  with the predictions of the hedgehog models.

Obviously something is wrong.   Two apparently model-independent
calculations of the same quantity give different results.  One
possibility, of course, is that one approach or the other is simply
wrong.   A more interesting possibility is that each of these
predictions is in fact correct in its domain of validity, but that the
domains of validity are mutually inconsistent.  Before exploring this
possibility in detail  it is worth
noting that there is a  hint that something strange is going on.
If one looks at the ratio of the hedgehog model predictions of the
leading chiral behavior to the conventional $\chi$PT behavior one sees
that this ratio is identical for all quantities with the same quantum
numbers.  For  any chirally divergent quantity $A$ with angular
momentum and isospin quantum numbers of $J$ and $I$ in Table 1.
\begin{eqnarray}
\frac{A^{I=J=0}_{\rm hedgehog}}{A^{I=J=0}_{\chi \rm PT}} & =  3 \nonumber\\  \\
\frac{A^{I=1, J=1}_{\rm hedgehog}}{A^{I=1,J=1}_{\chi \rm PT} } &= 
\frac{3}{2}   \label{univ}
\end{eqnarray}
This universal, but quantum number dependent, behavior suggests 
quite strongly that there is some deep connection between the 
hedgehog models and  $\chi$PT despite the fact that they do not agree.

How can it be that the domains of validity of the two approaches are
mutually incompatible?
The hedgehog  models were
based on working to leading order in the large-$N_c$ approximation
and then subsequently taking the chiral limit.  In contrast, the
 $\chi$PT predictions are based solely on the chiral limit.   Thus, the
possibility exists that for these chiral singular quantities, the
large-$N_c$ limit and the chiral limit do not commute.  Indeed this
possibility was recognized more than ten years ago by Adkins and Nappi
(1983),   who conjectured that this noncommutativity explained the
discrepancy between the Skyrme model and $\chi$PT predictions for the
isovector charge radius.  An interesting development in recent years is
 that  this conjecture  has been shown to be correct and, more
importantly, the dynamics underlying the noncommutativity has been shown
to arise from the behavior of the $\Delta$ resonance (Cohen and
Broniowski, 1992; Dashen,  Jenkins and Manohar, 1994; Cohen, 1995).

\section{BARYONS IN THE LARGE $N_c$ AND CHIRAL LIMITS}

To see why the $\Delta$ plays such a pivotal role for these
quantities in the large-$N_c$ limit, it is useful to go back
 to $\chi$PT and ask why it is that one pion loop graphs give the
leading chiral singularity 
for a generic quantity.    The quantities being studied can be thought
of as the response of the system when it is perturbed by some external
probe.  The response can be calculated via perturbation theory. The
chiral singularities
come from the fact that in this perturbation calculation, there are
small energy denominators associated with nucleon-plus-one-pion states. 
As $m_{\pi} \rightarrow 0$, the $\pi$-N states with arbitrarily low
relative momentum give  energy denominators which are arbitrarily
small.  If the phase space for such states is large enough these
small energy denominators   can cause divergences in the chiral
limit of $m_\pi = 0$.   Thus, the chiral singularities typically arise
from Feynman graphs such as diagram a of Fig. (\ref{loop}).

The  implicit assumption in derving the $\chi$ PT expressions in Table
1 is that the only states
which are arbitrarily low in energy  are the nucleon-plus-pion states. 
However, as one approaches the large-$N_c$
limit, this ceases to be true.  As $N_c$ becomes large,  the $\Delta $
becomes nearly degenerate with the nucleon.  In
the hedgehog models  Eq. (\ref{rot}) gives $M_\Delta - M_N =
3/{\cal{I}}$ where ${\cal{I}} \sim N_c$ so that
$M_\Delta - M_N \sim 1/N_c$.  Moreover, the fact that $M_\Delta - M_N \sim 1/N_c$
is  model independent;
it can be derived using the large-$N_c$ consistency relations (Jenkins,
1993).   Thus, the assumption that nucleon plus pion states
are the only low-lying excitations of the nucleon is wrong as one approaches the
large-$N_c$ limit.   

 In the context of the large-$N_c$ limit, we can now revisit  the
argument that the one pion loop contributions  dominate the chiral
singular quantities.  It is essentially the same argument
before; however,  in addition to graphs such as those in diagram a of 
Fig. \ref{loop} where there is a $\pi$-N intermediate state, there are
also  $\pi$-$\Delta$ intermediate states (diagram b).  
Since  the N and $\Delta$ are degenerate in the large $N_c$ limit,  the loops
containing a $\Delta$ also have anomalously small energy denominators
which can yield results which diverge as $m_\pi \rightarrow 0$.
Moreover in the large-$N_c$ limit, the $\pi$-N-$\Delta$ coupling is
completely determined from the  $\pi$-N-N  coupling (or equivalently
from $g_A$ and various Clebsch-Gordan factors). 
For large $N_c$, both the
nucleon and $\Delta$ are arbitrarily heavy and can be treated as static.
Since the pion couples to a static source  derivatively, the vertex is an operator 
with $I=1$ and $J=1$;  the relative strengths of the pion's coupling to various 
baryons is given by the fourth equation of eqs.~(\ref{mi}).

Now consider the contributions to some operator from diagrams a and b of
Fig. \ref{loop} in the large-$N_c$ limit. 
 The energy denominators  in the two diagrams  become identical as $N_c
\rightarrow \infty$ since $(M_\Delta -M_N) \rightarrow 0$.  The only
difference between diagrams a and b
is an overall factor coming from the vertices.  The net contributions
of these depend on the quantum numbers of the observables since these
determine how the various Clebsch-Gordan factors in the vertices are combined.
 Explicitly doing these sums one finds that the contributions
for an operator $A^{I,J}$ (with isospin and angular momentum quantum
numbers $I$ and $J$) from the $\pi$-N loops (as in diagram a. of 
Fig. \ref{loop}) and $\pi$-$\Delta$ loops (as in diagram b.) are
related as follows:
\begin{eqnarray}
A^{I=J=0}_{\pi-\Delta}  & = & 2 \, A^{I=J=0}_{\pi-\rm N} \nonumber \\
A^{I=J=1}_{\pi-\Delta}  & = & \frac{1}{2} \, A^{I=J=1}_{\pi-\rm N} \nonumber \\
A^{I=1, J=0}_{\pi-\Delta}  &= & -1  \, A^{I=1, J=0}_{\pi-\rm N} 
\label{ND} 
\end{eqnarray}
The large-$N_c$ limit of chirally singular properties includes
 both of these contributions.  Adding them together and noting that
the $\pi$-N contribution is the conventional $\chi$PT prediction for the
chirally singular part, one finds
\begin{eqnarray}
A^{I=J=0}_{{\rm Large} N_c}  & = & 3 \, A^{I=J=0}_{\chi \rm PT} \nonumber \\
A^{I=J=1}_{{\rm Large} N_c}  & = & \frac{3}{2} \, A^{I=J=1}_{\chi \rm
PT}  \label{ND2}
\end{eqnarray} 
Comparing Eqs. (\ref{ND2})  with Eqs. (\ref{univ}) we see immediately 
that the large-$N_c$ version of chiral perturbation theory predicts
precisely what is calculated in the hedgehog models.  

For $I=1, J=1$ operators one finds that contributions
 from diagram b cancel those from diagram a.  
Thus, for example, the ln($m_\pi)$ term in the isovector charge 
radius should vanish owing to such a cancellation (Cohen, 1995).  From
Table 1 we see
that the ln($m_\pi)$ term is not present in the  hedgehog model predictions.

The Skyrme  models reproduce the chirally
singular behavior of chiral perturbation theory, provided 
one uses the large-$N_c$ generalization of chiral perturbation
theory.  This is deeply satisfying from a purely theoretical
perspective.   At a practical level, however, there is still a problem
with these models.  The
preceding analysis was done in the large-$N_c$ limit, which implicitly
took the N-$\Delta$ mass splitting to be small compared to
all energies of order $N_c^0$.  Unfortunately,  $m_\pi$
is of order $N_c^0$ and in the real world, it is not much larger  than $M_\Delta
- M_N$.  Indeed it is a factor of two smaller.  The
preceding analysis is formally correct but it is of questionable utility.

The preceeding analysis can be generalized
to take into account the fact that the $M_\Delta - M_N \sim m_\pi$.
Formally, this corresponds to considering an expansion in which 
$1/N_c$ and $m_\pi/\Lambda$ are each taken as small, with $N_c m_\pi$
treated as ${\cal O}(1)$.  Analysis based on this generalized expansion 
is  large $N_c$ chiral perturbation theory.
In practice, this expansion implies that the leading
order contribution for chirally singular quantities again comes
from the sum of diagrams a and b in Fig.~(\ref{loop}).  However, the
$\Delta$ mass is not taken to be degenerate with the nucleon mass.  Rather, the
mass difference is kept to all orders.
 The diagrams  can be evaluated without difficulty;
 the results are  expressed most simply in terms of 
a universal function $S(d)$ where $d \equiv (M_\Delta - M_N)/m_\pi$
and $s(d)$ is given by
\begin{eqnarray}
s(d) & \equiv & \frac{4}{\pi} \, \frac{1}{\sqrt{1-d^2}} \, {\rm tan}^{-1}
 \left(\sqrt{\frac{1 - d}{1+d}} \right) \,\,\,\,\,\,\, {\rm for} \,\,\,
d<1 \nonumber \\
s(d) & \equiv & \frac{4}{\pi} \, \frac{1}{\sqrt{d^2-1}} \, {\rm
tanh}^{-1} \left(\sqrt{\frac{d-1}{d+1}} \right) \,\,\,\, {\rm for}
\,\,\, d>1 \label{s}
\end{eqnarray}
The leading order results for a number of quantities are given in Table 2.  

The function $s(d)$ has some interesting properties.  As $d
\rightarrow 0$ ({\it i.e.} large-$N_c$ limit), $s(d)
\rightarrow 1$.  If one replaces  $s(d)$ 
by unity for any of the quantities in Table 2 one  reproduces the analogous
Skyrme model results.   As $d
\rightarrow \infty$ ({\it i.e} the chiral limit),
$s(d)$ goes to zero as $2 \, {\rm ln} (d) (\pi d)^{-1}$.  As  the chiral
limit is approached, the $\Delta$ loop contributions become small
compared to the nucleon loop and one reproduces the naive $\chi$PT
results.  Thus, the large-$N_c$ $\chi$PT calculations smoothly
interpolate between the naive large-$N_c$ results of the hedgehog
models  and the results of conventional $\chi$PT. 

In the real world, $d \approx 2.1$ and $s(d) \approx .47$.  Thus
$s(d)$ is just about half way between the large-$N_c$ value of unity
and the chiral value of zero. To the extent one studies
properties for which $s(d)$ plays an important role (which includes all
leading chirally nonanalytic properties),   low order expansions 
around either the chiral limit or the large-$N_c$ limit will be
incorrect.  We are far from both.

\section{IMPLICATIONS FOR MODELING BARYONS}

What does the preceding analysis tell us about modeling baryons?
Clearly one should use great care when interpreting the results of
the standard Skyrme model (or other hedgehog model) calculations for
quantities in which the long-distance part of the pion tail is
important.   Errors due to the  large-$N_c$ limit 
 (upon which calculations in the model are based) are likely to be very large
since in these calculations $s(d)$ is implicitly taken to be unity.  What
alternatives are there?

One possibility is simply to declare the long-range chiral physics
uninteresting and beyond the scope of the model.  The hope is 
that by suitable choice of phenomenological parameters one can simulate
the correct long-distance physics without including the  explicit
pionic degrees of freedom.  This is the strategy underlying the
constituent quark model.   On the other hand, the long distance chiral
physics is one of the few nonpertubative aspects of QCD that we
actually understand.  If we hope to learn something fundamental 
from the models about how QCD plays out in the low energy regime,
we should at least be able to explain these simple features.  In this
context, it is also worth remarking that the relationship of the
constituent quark model to QCD remains quite cloudy.

Another alternative is to simply abandon the enterprise of modeling
entirely.  One can  develop large-$N_c$ chiral perturbation theory in a
systematic and model-independent way.  At any given order in the expansion, one
will have only a finite number of parameters, which should be fitted
experimentally.  Having fit these parameters, one can then proceed to
predict other quantities.  Initial steps in this program have been
taken ,and the results are promising (see, for example, Luty and
March-Russell (1995); Jenkins and Lebed (1995);  Bedaque and Luty
(1995); Jenkins (1995)).  However,
this approach has at least one important limitation:   the regime of
validity of the expansion is unknown.  It is
not known how high in momentum transfer the theory will give useful
results.

A final approach would be to consider models which formally give both
the large-$N_c$ and chiral limits correctly, and which does not
implicitly assume  that0 $M_\Delta -M_N$ is either much smaller or
much larger than $m_\pi$.  One class of models which does this includes
the cloudy bag model (for a review see Thomas (1983)), which has a
quark core to which one adds explicit quantum pion loops.  Because
the pions are treated quantum mechanically from the outset, 
they automatically get the full function $s(d)$  as $N_c \rightarrow \infty$
and $m_\pi \rightarrow 0$ .  One can enhance the constituent quark
model in a similar way by including explicit pion loops.

 Another possibility is to develop a new calculational scheme for studying
 the Skyrme model and
its hedgehog cousins.  The analysis of the Skyrme model, which 
showed inconsistencies with the chiral behavior, was based on the conventional
calculation scheme, which implies a 
$1/N_c$ expansion from the outset.  If one could consistently 
reformulate the calculation of quantities in the Skyrme model in a fashion
consistent with the expansion scheme of large $N_c$ chiral perturbation theory
there would be no difficulty.  Is there a
calculational scheme consistent with an expansion?  The answer,
apparently, is that there is. 
 This scheme was recently introduced by Dorey, Hughes and Mattis (1994). It has been
christened the ``rotationally improved skyrmion'' (RISKY) and is  based
on a functional integral formulation around a hedgehog distorted by its
rotational motion.  One test of the consistency of this approach is its
ability to reproduce the physics associated with the ratio of the $\Delta$
nucleon mass splitting to $m_\pi$.  Using RISKY techniques, the
functional dependence on this ratio, {\it i.e.},
$s(d)$, was reproduced for one of the quantities in Table 2
(Dorey, Hughes and Mattis, 1995), and presumably the appropriate $s(d)$
factors for other quantities can also be obtained.   A major drawback of
the RISKY approach is that it is quite cumbersome.  Indeed, at present,
the RISKY approach has not yet  been implemented fully for any particular Skyrme
model Lagrangian in the sense that the various low energy observables have not been 
computed in terms of the parameters in the Lagrangian.

The general question of how to model the low energy properties
of baryons remains an interesting and vibrant field of study.  
The connection of these models to QCD is clearly a central
issue.   In this context, it is important that good models should, at a
minimum, reproduce those  aspects of QCD which we actually understand. 
Fortunately, in the past several years
our understanding of certain nonperturbative aspects of QCD has
increased; particularly the large-$N_c$ and chiral properties of
baryons.   Serious models should build upon this understanding.

\section{ ACKNOWLEDGMENTS}

This work was completed in part at the Institute for Nuclear Theory (INT) of  the
University of Washington.  The hospitality of the INT is gratefully
acknowledged.  The support of the U. S. Department of Energy (grant
no.  DE-FG02-93ER-40762) and the U.  S. National Science Foundation
(grant no. PHY-9058487) is alsogratefully acknowledged.

\appendix

\section{CHIRAL PROPERTIES IN SKYRME MODELS}

In this appendix, it is shown how the ``model-independent'' predictions
of chirally nonanalytic  quantities are calculated  in Skyrme and other
large-$N_c$ hedgehog models.  The basic strategy is to use
the standard semiclassical treatment (Adkins, Nappi, and Witten 1983;
Cohen and Broniowski, 1986).  In this approach,
the various operators are treated as classical with the classical
fields from the hedgehog configuration.  The only quantum mechanics 
come from the quantization of the collective 
rotational (or, equivalently, isorotational) modes.
The important thing to note is that if a quantity diverges in the chiral
limit, it implies that the quantity is dominated by the long-range
pionic tail, whose range goes to infinity as $m_\pi \rightarrow 0$.
Thus for the chirally divergent part of these quantities, it is
sufficient to pick up the contributions due to the long-range part of the pions.

Consider as a concrete example the isovector charge radius. 
The pion fields in an arbitrary hedgehog
 configuration may be written as
$\pi_a(\vec{x}) = \pi(r)\hat{r}_a$, where $a$
 is the isospin direction.  In any of these hedgehog models  the
 pion field tends asymptotically  
to a p-wave Yukawa form with a field
 strength fixed by  $g_{\pi N N}$ as given by eq.~(\ref{yuk})
where $g_{\pi N N}$ is the model's $\pi$-N coupling.  
 Following Cohen and Broniowski (1986), the asymptotic pion field
contribution to the  isovector charge radius is given by
\begin{equation}
\langle r^2 \rangle_{I=1} \, = \,
 \frac{1}{\cal I} \, \int \, {\rm d}r \, r^4
 \, \frac{8 \pi}{3} ({(\pi_{\rm asympt})}^2  \,\,\, ,
\label{cr}
\end{equation}
where ${\cal I}$ is the
 moment of inertia.   Other contributions to $\langle r^2 \rangle_{I=1}$
  come from shorter-ranged   effects, which are model-dependent and 
do not diverge in the chiral limit.
 The expression in Eq. ~(\ref{cr})  is easy to
understand--it is the asymptotic pion field contribution to 
the semiclassical isovector-vector current multiplied by $r^2$.
Evaluating the
integral and using the relation of 
 Goldberger and Treiman (1958),
 $g_{\pi N N} f_{\pi} = g_{A}  M_N$ (which
 is true to leading order in the pion mass)
 and using the fact that for hedgehog models $1/{\cal I} = 2/3
 (M_\Delta -M_N)$ (to leading order in the
 $1/N_c$ expansion), immediately yields
\begin{equation}
 \langle {\rm N}| r^2 |{\rm N} \rangle_{I=1}
 \, = \, \frac{5 \, g_A^2 \,  \delta m}{16 \, \pi \, f_\pi^2 \, 
m_\pi} 
\end{equation}
plus corrections which are
 higher order in either $1/N_c$ or $m_q$. 

Model-independent relations for other quantities in Table 1
can be  derived in an analogous fashion.

\section{CONSISTENCY AND LARGE $N_c$}

This appendix briefly discusses the physical basis for the
large-$N_c$
consistency relations originally derived by  Gervais and Sakita (1984a,1984b)
and subsequently rediscovered and extended by
 Dashen and Manohar (1994).  There has been considerable work on this
subject in the past several years (Jenkins 1993; Dashen,  Jenkins and 
Manohar, 1994,1995; Broniowski, 1994). 
In this appendix, mathematical detail will be omitted for reasons of space.  

The basic argument is quite simple.  Consider $\pi$-N scattering.  From
Witten's large-$N_c$ analysis of generic baryons (Witten, 1979), we
expect that the effective $\pi$-N coupling for derivatively coupled
pions, $g_{\pi N N}/M_N = g_A/f_\pi$, is of order $N_c$.  This implies that
the Born and crossed-Born contributions in Fig. 1 are of order
$N_c^2$.   Moreover, the non-Born terms have a different energy
dependence owing to the presence of a mass difference in the propagator
and cannot cancel out the Born and crossed-Born contributions.  This
presents a serious problem: as $N_c \rightarrow \infty$, the Born and
crossed-Born contributions become arbitrarily large and, unless they
cancel each other, violate unitarity.  

It is easy to see that these contributions might cancel.  In the large
$N_c$ limit, the nucleon is very massive and does not recoil.  Thus, the
energy denominator in the Born and crossed-Born terms are
equal and opposite.  Indeed, for scalar or isoscalar mesons, this fact
alone is sufficient to guarantee cancelation.  However, the pion
coupling is a p-wave isovector and thus, in the nonrelativistic
limit--which is appropriate at large $N_c$--goes as $N^\dagger 
\sigma_i \tau_a N \partial_i \pi^a$.  Thus the sum of the Born and
crossed-Born terms is proportional to $N_c^2 \, [\sigma_i \tau_a, \sigma_j
\tau_b]$, which is nonzero.  How then can one have consistency between
unitarity and Witten's large-$N_c$ prediction for the scaling of the
$\pi$-N coupling constant?  

The answer is straightforward.  If the $\Delta$ were degenerate with
the nucleon in the large-$N_c$ limit, then the $\Delta$ graphs could
cancel the nucleon Born graphs. 
Taking into consideration $\pi$-$\Delta$
scattering, one deduces the need for an additional $I=J=5/2$ state to
keep the amplitude unitary.  The  existence of
the entire tower of states that is found in the Skyrme model is deduced.
If one
uses a vector-isovector operator, $X_{i,a}$,
to  describe all of the $\pi$-$B$-$B^{\prime}$ couplings, where $B$ and
$B^\prime$ are baryons in the tower, then the consistency condition implies
\begin{equation}
[X_{i,a},X_{j,b}] = 0  \; \; .
\end{equation}
{From} this equation, a  simple recursion relation follows and allows
the computation of matrix elements of $X$ in terms of Clebsch-Gordan coefficients.  

To obtain other consistency relations one can apply similar reasoning
to scattering amplitudes for $\pi \, + \ B \, \rightarrow 
\pi \, + \, m + B$,  where $m$ is a meson with fixed quantum numbers.
In the special case where $m$ is a pion, one finds that 
model-independent relations hold to relative order $1/N_c^2$.  One can 
apply the same technique to the reaction $\pi \, + \ B \, \rightarrow 
\pi \, + \, \gamma + B$,  which constrains electromagnetic coupling,
and to processes which involve weak interactions.\\

\vspace{.18in}
\noindent REFERENCES
\vspace{.18in}

\noindent Adkins,  G. S. , C. R. Nappi, 1984, Nucl. Phys.  {\bf B 233},  109.

\noindent Adkins,  G. S. , C. R. Nappi, 1985, Nucl. Phys.  {\bf B 249},  507.

\noindent Adkins,  G. S. , C. R. Nappi,
 and E. Witten, 1983, Nucl. Phys.  {\bf B 228},  552. 

\noindent Alkofer, R., H. Reinhardt and H. Weigel, 1995,  ``Baryons as Chiral
solitons in the Nambu--Jona-Lasinio model'', to appear in Phys. Rep.
{\bf 263};  hep-ph/9501213.

\noindent Banerjee, M. K., W. Broniowski, and T. D. Cohen, 1987,
 in {\em {Chiral Solitons}},
 edited by K.-F. Liu (World Scientific,
  Singapore), p. 255. 

\noindent Bernard, V.,  N. Kaiser and U.-G. Meissner, 1995, Int. J. Mod. Phys.
{\bf E 4}, 193. 
 
\noindent Birse, M. C., 1991, 
 Prog. in Part. and Nucl. Physics 
  {\bf 25}, 1. 

\noindent Bedaque, P. F., and M. Luty, 1995, ``Baryon masses at second order in
large-$N$ chiral perturbation theory'',
hep-ph/9510453.  

\noindent Broniowski, W., 1994,   Nucl. Phys. {\bf A 580}, 429. 

\noindent Chodos, A.,  R. L. Jaffe, K. Johnson, C. B. Thorn, V.F.
Weisskopf, 1974,  Phys. Rev. {\bf D 9} 3471.

\noindent Close, F., 1979, {\it An Introduction To Quarks and Partons},
(Academic Press, London).

\noindent Cohen, T. D, 1995, Phys. Lett. {\bf B 359}, 23.              

\noindent Cohen, T. D.  and W. Broniowski, 1986,
 Phys. Rev. {\bf D 34}, 3472.

\noindent Cohen, T. D.  and W. Broniowski, 1992,
Phys. Lett.  {\bf B 292},  5.

\noindent Dashen, R.,  E. Jenkins and A. V. Manohar, 1994,  
 Phys. Rev. {\bf D 49}, 4713.

\noindent Dashen R., E. Jenkins and A. V. Manohar, 1995,  
 Phys. Rev. {\bf D 51}, 2489.

\noindent Dashen R. and A. V. Manohar, 1993, Phys.
 Lett.{\bf  B 315},   42.
 
\noindent  Dorey, N.,  J. Hughes, M. P. Mattis, 1994, Phys. Rev. {\bf D 50}, 5816.

\noindent Dorey, N.,  J. Hughes, M. P. Mattis and D. Skidmore, 1995,
``Chiral divergent properties of hadrons in the large-$N_c$ limit''
Michigan State Preprint MSU-50605;  hep-ph/9509310.  

\noindent Gell-Mann, M., R. J. Oakes and B. Renner, 1968, 
Phys. Rev. {\bf 175}, 2195.

\noindent Gervais, J. L  and B. Sakita, 1984a, Phys. 
Rev. Lett. {\bf 52},  87.
 
\noindent Gervais, J. L  and B. Sakita, 1984b, 
Phys. Rev. {\bf D 30}, 1795.

\noindent Goldberger, M.  and S. Treiman, 1958, 
 Phys. Rev.  {\bf 111},  354.

\noindent Isgur, N., 1992, in {\it Nucleon Resonances and Nucleon Structure}, 
editied by G. A. Miller (World Scientific, Singapore), p. 45. 
 
\noindent  Jenkins, E., 1993,  Phys. Lett. {\bf B 315},  441.

\noindent  Jenkins, E. and R. F. Lebed, 1995, Phys. Rev. {\bf D 52},  282.
\noindent  Jenkins, E., 1995, ``Chiral Lagrangian for Baryons in the
$1/N_c$ Expansion'' , University of California San Diego preprint 
UCSD/PTH 95-17; hep-ph/9509433

\noindent  Karl, G., 1992, in {\it Nucleon Resonances and Nucleon Structure}, 
edited by G. A. Miller (World Scientific, Singapore), p. 71. 

\noindent Karsch, F., J. Engels, E. Laermann, B. Petersson, 1995, Eds.
{\it Lattice '94. Proceedings , 12th International Symposium On Lattice
Field  Theory} (North-Holland, Amsterdam, Netherlands) (Nucl. Phys.
B, Proc. Suppl. 42).

\noindent Luty, M. A.  and J. March-Russell, 1995, Nucl. Phys, {\bf B 426}, 2322.

\noindent Skyrme, T. H. R.,  1961a, Proc. R. Soc. London  {\bf 260}, 126.

\noindent Skyrme, T. H. R.,  1961b, Proc. R. Soc. London  {\bf 261}, 237.

\noindent Skyrme, T. H. R.,  1962, Nuc. Phys,{ \bf 31}, 556.

\noindent Thomas, A. W., 1983, {\it Advances in Nuclear Physics} {\bf
13}, edited by J. Negele and E. Vogt (Plenum, New York), p. 1.

\noindent Witten, E., 1979,  Nuc. Phys {\bf B 160}, 57.

\noindent Vepstas, L.  and A. D. Jackson, 1990, 
Phys. Rep.  {\bf 187},  109.
 
\noindent  Zahed, I and  G. E. Brown, 1986, Phys.
Rept.  {\bf 142}, 1. \\

\clearpage

\newpage
\begin{table}
\noindent Table 1: The leading chiral nonanalytic contribution for  a number of
nucleon properties.  These contributions will dominate the full answer
in the limit $m_\pi \rightarrow 0$.   The column headed ``hedgehog
models'' gives the ``model-independent'' predictions for the Skyrme
model or any other large-$N_c$ hedgehog model.  The column headed
``Naive $\chi$PT'' gives the prediction from chiral perturbation
theory based on the assumption that the pion is the only light
excitation in the problem.  All of these quantities are experimentally
observable except for those labeled with a ${}^\dagger$.  These
quantities concern the quark mass dependence of observables; for
convenience these quantities are re-expressed in terms of pion mass
dependences.   In principle these quantities could be extracted from
numerical simulations of QCD. 
$$
\begin{array}{|| c | c |  c | c ||} \hline \hline 
{\rm Quantity} & {\rm Quantum} \#'s & {\rm Hedgehog \, \, \, Models } &
{\rm Naive} \, \, \chi {\rm PT} \\ \hline
& & & \\
{\rm scalar \, \,  radius}  & I=0 & & \\
\langle r^2 \rangle_s \, = \, 6 \,
\frac{\partial~\sigma(t)}{\partial~t} |_{t=0} & J=0 &
\frac{g_a^2}{f_\pi^2} \frac{9 m_\pi}{64 \pi}  & 
\frac{g_a^2}{f_\pi^2} \frac{3 m_\pi}{64 \pi} \\  \hline
& & & \\
{\rm quark\, \, mass \, \,dependence}  & &  &  \\
{\rm of} \; M_N  \; {\rm  (in \; terms \; of }\, m_\pi^2{\rm )} & I=0 &
 - \frac{g_a^2}{f_\pi^2} \frac{27}{128 \pi m_\pi} & -
\frac{g_a^2}{f_\pi^2} \frac{27}{128 \pi m_\pi} \\
\frac{\partial ^2 M_N}{\partial (m_\pi^2)^2} \, \, \, \,{}^\dagger & J=0
& &  \\ \hline
& & & \\
{\rm electric \, \, polarizability} & & & \\
{\rm of \, nucleon} & I=0 &  \frac{g_a^2}{f_\pi^2} \frac{5 e^2}{128
\pi^2 m_\pi} & \frac{g_a^2}{f_\pi^2} \frac{5 e^2}{384 \pi^2 m_\pi} \\ 
\alpha_N & J=0 & &\\ \hline
& & &\\
{\rm magnetic \, polarizability} & & & \\
{\rm of \, the \, nucleon} & I=0 &  \frac{g_a^2}{f_\pi^2} \frac{ e^2}{256
\pi^2 m_\pi} & \frac{g_a^2}{f_\pi^2} \frac{ e^2}{768 \pi^2 m_\pi} \\
\beta_N & J=0 & & \\ \hline
& & & \\
{\rm isovector \, \,  magnet} & & & \\
{\rm radius  \, of \, nucleon}  & I=1 & \frac{1}{\mu^{\rm anom}_{I=1}}
\frac{g_a^2}{f_\pi^2} \frac{3}{8 \pi m_\pi} &  {\mu^{\rm anom}_{I=1}}
\frac{g_a^2}{f_\pi^2} \frac{1}{4 \pi m_\pi} \\
\langle r^2 \rangle^{I=1}_{\rm mag} = \frac{6}{\kappa_{I=1}}
\frac{\partial F_2(t)}{\partial t}  |_{t=0} & I=1 & & \\ \hline
& & & \\
{\rm quark\, \, mass \, \,dependence }  & &  &  \\
{\rm  of \, \, anomalous \, \,  isovector } &  &   &  \\
{\rm moment \, \,   (in \, \, terms \, \,of } m_\pi^2{\rm ) } & I=1 & 
\frac{g_a^2}{f_\pi^2} \frac{3}{8 \pi m_\pi}& \frac{g_a^2}{f_\pi^2}
\frac{1}{4 \pi m_\pi}\\
\frac{\partial \mu^{\rm anom}_{I=1}}{\partial m_\pi^2} \, \, \, \, 
{}^{\dagger} & J=1 & & \\ \hline
& & & \\
{\rm nucleon \, \,  isovector } &  & & \\
{\rm charge radius} & I=1 & \frac{g_a^2}{f_\pi^2} \frac{M_\Delta -
M_N}{m_\pi}  \frac{5}{16 \pi} & \frac{5g_a^2 +1 }{f_\pi^2} \frac{{\rm
ln}(m_\pi)}{8 \pi^2}\\
\langle r^2 \rangle^{I=1}_{\rm elec} = 6 \frac{\partial
F_1(t)}{\partial t}  |_{t=0} & J=0 & & \\
 & & & \\ \hline \hline
\end{array}
$$
\end{table}

\clearpage

\newpage

\begin{table}
\noindent Table 2: The leading chiral nonanalytic contribution for a number of
nucleon properties in large-$N_c$ chiral perturbation theory.  These
contributions will dominate the full answer in the limit $m_\pi
\rightarrow 0, \; \, N_c \rightarrow \infty, \; \, N_c m_\pi
\rightarrow $ constant.  The function $s(d)$ is given in Eq.
(\ref{s}); $d \equiv (M_\Delta-M_N)/m_\pi$.  All of these quantities
are experimentally observable directly except for those labeled with a
${}^\dagger $.

$$
\begin{array}{|| c | c | c ||} \hline \hline 
{\rm Quantity} & {\rm Quantum} \#'s &  {\rm Large} \, \, N_c  \chi
{\rm PT} \\ \hline
& & \\
{\rm scalar \, \,  radius}  & I=0 & \\
\langle r^2 \rangle_s \, = \, 6 {\frac{\partial \sigma(t)}{\partial t}}
|_{t=0} & J=0 & (1 \, + \, 2 \, s(d) \, ) \,  \frac{g_a^2}{f_\pi^2}
\frac{3 m_\pi}{64 \pi} \\ \hline
& &  \\
{\rm quark\, \, mass \, \,dependence}   &  &  \\
{\rm of} \, \, M_N \, \, {\rm  (in \, \, terms \, \,of }m_\pi^2{\rm )}
& I=0 &  - (1 \, + \, 2 \, s(d) \, ) \,  \frac{g_a^2}{f_\pi^2}
\frac{27}{128 \pi m_\pi} \\
\frac{\partial ^2 M_N}{\partial (m_\pi^2)^2} \, \, \, \,{}^\dagger &
J=0 &  \\ \hline
 & & \\
{\rm electric \, \, polarizability}  & & \\
{\rm of \, nucleon} & I=0  & (1 \, + \, 2 \, s(d) \, ) \,
\frac{g_a^2}{f_\pi^2} \frac{5 e^2}{384 \pi^2 m_\pi} \\ 
\alpha_N & J=0  & \\ \hline
& & \\
{\rm magnetic \, polarizability}  & & \\
{\rm of \, the \, nucleon} & I=0  & (1 \, + \, 2 \, s(d) \, ) \,
\frac{g_a^2}{f_\pi^2} \frac{ e^2}{768 \pi^2 m_\pi} \\
\beta_N & J=0 &  \\ \hline
& & \\
{\rm isovector \, \,  magnet} & & \\
{\rm radius  \, of \, nucleon}  & I=1 &  (1 \,  + \,  \frac{1}{2} s(d)
\, ) \,  {\mu^{\rm anom}_{I=1}} \frac{g_a^2}{f_\pi^2} \frac{1}{4 \pi m_\pi} \\
\langle r^2 \rangle^{I=1}_{\rm mag} = \frac{6}{\kappa_{I=1}}
\frac{\partial F_2(t)}{\partial t}  |_{t=0} & I=1 &  \\ \hline
& &  \\
{\rm quark\, \, mass \, \,dependence }  &   &    \\
{\rm  of \, \, anomalous \, \,  isovector } &  &     \\
{\rm moment \, \,   (in \, \, terms \, \, of } m_\pi^2{\rm ) } & I=1 & (1
\,  + \,  \frac{1}{2} s(d) \, ) \, \frac{g_a^2}{f_\pi^2} \frac{1}{4 \pi m_\pi}\\
\frac{\partial \mu^{\rm anom}_{I=1}}{\partial m_\pi^2} \, \, \, \,  {}^\dagger & 
J=1 &  \\ \hline
& &  \\
{\rm nucleon \, \,isovector }   & & \\
{\rm charge \, radius} & I=1 & d \, s(d)\, \frac{g_a^2}{f_\pi^2} \frac{5}{16 \pi}\\
\langle r^2 \rangle^{I=1}_{\rm elec} = 6 \frac{\partial
F_1(t)}{\partial t} |_{t=0} & J=0 & \\  \hline \hline
\end{array}
$$
\end{table}

\clearpage

\newpage

\begin{figure}[th]
\caption{\noindent Typical pion loop  diagrams which contribute to the leading
chiral nonanalytic behavior for nucleon properties  in the
large-$N_c$
limit of QCD.  The cross represents the particular external source
being probed--electric, magnetic, scalar, {\it etc.}  Diagram a.~gives
the leading nonanalytic behavior as one approaches  the pure chiral
limit ($m_{\pi} \ll (M_\Delta -M_N)$).  Diagram b.~becomes comparable
when $m_{\pi} \approx (M_\Delta -M_N)$.}
 \label{loop}
\end{figure}

\begin{figure}[bh]

\end{figure}

\noindent Figure 2:  Born and crossed-Born contributions to $\pi$-N scattering.

\clearpage

\end{document}